\documentclass[conference, a4paper]{IEEEtran}
\usepackage[left=1.32cm,right=1.32cm,top=1.9cm,bottom=4.3cm]{geometry}
\usepackage{subcaption}
\usepackage{cite}
\usepackage{amsmath,amssymb,amsfonts}
\usepackage{graphicx}
\usepackage{textcomp}
\usepackage{tikz}
\usepackage{adjustbox}
\usepackage{multirow}
\usepackage{multicol}
\usepackage[affil-it]{authblk} % Pour les affiliations

\usepackage{xcolor}
\usepackage{pgfplots}
\pgfplotsset{compat=newest,compat/show suggested version=false}
\usepackage{booktabs}
\usepackage{algpseudocode}
\usepackage[ruled, lined, longend, linesnumbered]{algorithm2e}
\usepackage[utf8]{inputenc}
\usepackage{tikz}
 
\usepackage{amssymb}% http://ctan.org/pkg/amssymb
\usepackage{pifont}% http://ctan.org/pkg/pifont
\def\BibTeX{{\rm B\kern-.05em{\sc i\kern-.025em b}\kern-.08em
    T\kern-.1667em\lower.7ex\hbox{E}\kern-.125emX}}
\begin{document}
% \verb|\checkmark|: \checkmark \par
% \verb|\cmark|: \cmark \par
% \verb|\xmark|: \xmark
\title{Beyond Detection: Leveraging Large Language Models for Cyber Attack Prediction in IoT Networks}

%\title{Large Language Models-based Intrusion Prediction for IoT Networks}

%\title{From Detection to Prediction: Leveraging Large Language Models for Dynamic Cyber Attack Projection in IoT Networks}

\author[1]{Alaeddine Diaf}
\author[1,3]{Abdelaziz Amara Korba}
\author[2]{Nour Elislem Karabadji}
\author[3]{Yacine Ghamri-Doudane}

\affil[1]{LRS, Badji Mokhtar Annaba University, Algeria}
\affil[2]{National Higher School of Technology and Engineering, LTSE, Algeria}
\affil[3]{L3I, University of La Rochelle, France}

\maketitle
\begin{abstract}
%The large-scale adoption of Internet of Things (IoT) technology has led to an increase in the number and severity of cyberattacks.  Despite real-time IDSs, attacks still happen. Accordingly, there is a need to predict attacks before they occur in order to avoid catastrophic damage. This paper introduces an intrusion prediction system based on Large Language Models and Long Short Term Memory (LSTM), which capable of predicting intrusions based on network packets. The proposed intrusion prediction system harnesses the potential of two transformer-based Pre-trained Large Language Models (LLMs), GPT and BERT, for predicting and evaluating next network packets. A fine-tuned GPT functions as the network packet predictor, while a fine-tuned BERT serves as an evaluator for GPT's predictions. Additionally, a trained LSTM model is employed for classifying predicted packets and detecting malicious ones. Our extensive experiments on the recently published IoT attack dataset, CICIoT2023, shows that the proposed system achieved a high intrusion prediction rate with an overall accuracy of 98\%.

In recent years, numerous large-scale cyberattacks have exploited Internet of Things (IoT) devices, a phenomenon that is expected to escalate with the continuing proliferation of IoT technology. Despite considerable efforts in attack detection, intrusion detection systems remain mostly reactive, responding to specific patterns or observed anomalies. This work proposes a proactive approach to anticipate and mitigate malicious activities before they cause damage. This paper proposes a novel network intrusion prediction framework that combines Large Language Models (LLMs) with Long Short Term Memory (LSTM) networks. The framework incorporates two LLMs in a feedback loop: a fine-tuned Generative Pre-trained Transformer (GPT) model for predicting network traffic and a fine-tuned Bidirectional Encoder Representations from Transformers (BERT) for evaluating the predicted traffic. The LSTM classifier model then identifies malicious packets among these predictions. Our framework, evaluated on the CICIoT2023 IoT attack dataset, demonstrates a significant improvement in predictive capabilities, achieving an overall accuracy of 98\%, offering a robust solution to IoT cybersecurity challenges.
\end{abstract}


\section{Introduction}
In today's interconnected world, the Internet of Things (IoT) plays a pivotal role in shaping and optimizing various aspects of our daily lives, revolutionizing the way devices, systems, and data seamlessly interact. The large-scale adoption of IoT devices across diverse fields has resulted in the emergence of novel cyberattacks specifically designed for IoT, with the intent of exploiting the vulnerabilities present in these interconnected devices. Sophisticated cyberattacks, allows unauthorized or malicious activities that compromise the security of IoT network which is referred as a network intrusion \cite{abdulganiyu2023systematic}. Network intrusions can take various forms, including unauthorized access to sensitive information, denial-of-service attacks, or the introduction of malware into the network. Detecting and preventing network intrusions are critical aspects of network security to ensure the integrity, confidentiality, and availability of network resources. To maintain these security requirements, various techniques have been implemented to counter intrusions in IoT networks. Among the existing techniques, Intrusion Detection Systems (IDSs) play a crucial role in identifying potential security threats within IoT networks. In recent years, Artificial Intelligence-based IDSs have gained significant attention from the research community due to their capability to achieve real-time intrusion detection. Nevertheless, their efficacy in anticipating and preemptively mitigating malicious events before they occur is constrained, they are often reactive in nature
\cite{heidari2023internet,saied2024review,asharf2020review}.  Given this limitation of intrusion detection systems, it is essential to prioritize intrusion prediction for enhancing the strength and effectiveness of cybersecurity measures. Intrusion prediction enhances a security posture by introducing a forward-looking dimension, enabling organizations to stay one step ahead of cyber adversaries and significantly reducing the likelihood and impact of successful intrusions.  In this context, using Pre-trained Large Language Models (LLMs) presents a cutting-edge approach in the field of cybersecurity. Adapting LLMs for various threat detection has been studied \cite{Liu2024,Seyyar2022,Chen2022,Ferrag2023,Ferrag2023SecureFalcon,han2023loggpt}. Leveraging the inherent linguistic capabilities of these models, such as Generative Pre-trained Transformer (GPT)\cite{radford2019language} and Bidirectional Encoder Representations from the Transformers (BERT) \cite{devlin2019bert} models, for analyzing network patterns and anomalies opens new avenues for proactive threat detection. By harnessing the contextual understanding and pattern recognition abilities embedded in LLMs, intrusion prediction systems can be enhanced, providing a more robust defense against evolving cyber threats.
In this paper, we present a novel intrusion prediction framework based on network packets, employing a combination of Fine-tuned Pre-trained LLMs and LSTM model. The proposed framework is designed to predict potential intrusions in IoT networks by predicting next network packets giving current ones using a generative pre-trained LLM and classifying them through the LSTM model. The assessment of the predicted network packets leverages the bidirectional contextual understanding provided by the BERT model.  This framework is required to grasp the fundamental features of network packets, accurately predict their subsequent packets, and efficiently predict intrusions. As far as we know, this is the first attempt to propose a pretrained large language models for network intrusion prediction. We performed fine-tuning of GPT on both normal and malicious network traffic , aiming to predict the next network packets. Additionaly, BERT was fine-tuned for a packet-pair classification task to assess the predicted packets from the fine-tuned GPT. Furthermore, predicting intrusions using a trained LSTM model. We evaluate the effectiveness of our proposed framework using a recent IoT attack dataset. Based on our experimental analysis, our approach successfully detects intrusions with an impressive overall accuracy of 98\%.\par
The remainder of this paper is organized as follows. Section II presents related work. Section III  delves into the methodology of the proposed framework. Section IV depicts the performance evaluation results and finally, Section V concludes the paper.

\section{Related Work} \label{RT}

Several recent LLMs-based IDSs have been proposed for detecting attacks in IoT networks. BERT is widely fine-tuned for this purpose. The authors of \cite{Liu2024} proposed SeMalBERT model for identifying malicious software in Windows systems. Where BERT was fine-tuned to learn behavioral features of API call sequences for semantic-based malware detection, enabling more detailed and context-aware malware detection. However, this model is complex faces challenges related to computational resources. In \cite{Chen2022}, BERT’s bidirectional contextual understanding has been leveraged for effectively identifying anomalies in system log data, it shows an impressive F1-score of 99.3\% in anomalies detection. A similar approach to detect anomalous HTTP requests was proposed in \cite{Seyyar2022}. In another effort, the authors in \cite{Wang2024} introduces a lightweight intrusion detection model tailored for IoT, leveraging an enhanced version of BERT-of-Theseus. This model aims to enhance the efficiency of intrusion detection in the context of IoT devices, presenting a notable advancement in lightweight security solutions. Other LLMs have also been fine-tuned for cybersecurity aims, a fine-tuned version of Falcon \cite{penedo2024refinedweb} was proposed in \cite{Ferrag2023SecureFalcon}, called SecureFalcon to identify vulnerabilities in C codes. Additionally, GPT has been fine-tuned to develop VulDetect \cite{omar2023vuldetect}, a vulnerability detection framework. With an accuracy rate of up to 92.65\%, VulDetect effectively identifies software vulnerabilities. 

Overall, existing LLMs-based cybersecurity solutions are focusing only on intrusion detection, aiming at preventing known attacks and stopping ongoing threats. However, Stopping multistage attacks in its earlier stages and predicting the ultimate attack to avoid its catastrophic damages, have been ignored.  Considering these gaps, we propose a novel network packet-based intrusion prediction framework designed using LLMs and LSTM model. Our framework can predict intrusions based on current network packets, predicting  multistage and unseen attacks.

\section{Proposed Solution}\label{SOL}

%justification du chooix (gpt, bert and lstm) and how they work collaboratively

This section outlines the crucial steps of our proposed framework, consisting of three essential elements: packet parsing and preparing, fine-tuning pre-trained LLMs for predicting the next packets, and training LSTM to classify network packets.

\begin{figure*}
    \centering
    \includegraphics[scale=0.4]{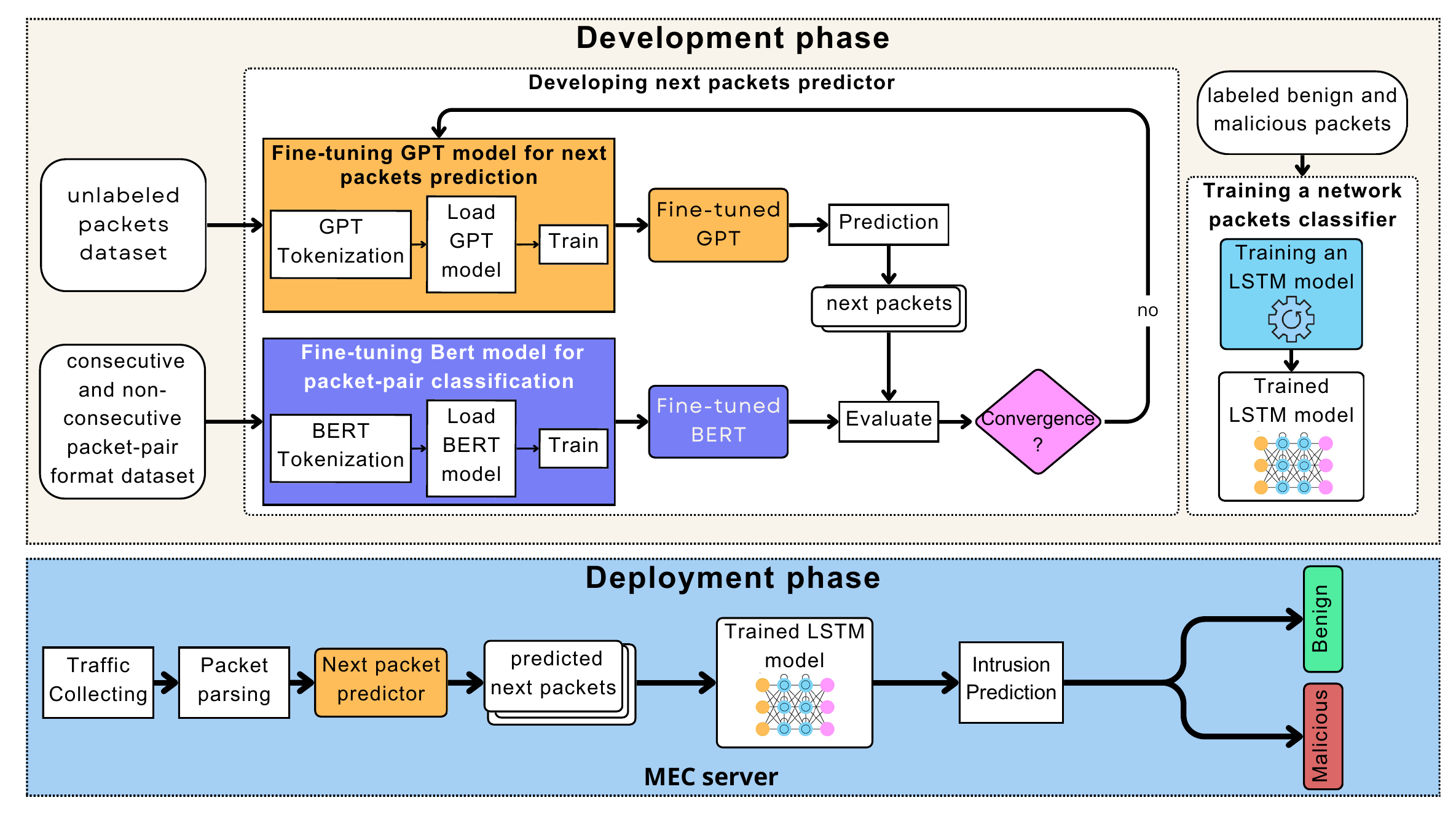}
    \caption{Workflow of the proposed intrusion prediction framework }
    \label{fig:archi3}
\end{figure*}

During the development phase of our intrusion prediction framework (see figure \ref{fig:archi3}), we pass through three key elements: fine-tuning GPT for next packets prediction, fine-tuning BERT to asses next packets prediction and training an LSTM packet classifier.

Regarding temporal dependence between consecutive packets, the network packets transmitted within an IoT network can be represented as sequence or a time series P = \{$p^{(1)}$, $p^{(2)}$,..., $p^{(n)}$\}, where each packet $p^{(t)}$ is an m-dimensional vector \{$f_{1}^{(t)}$, $p_{2}^{(t)}$,..., $p_{m}^{(t)}$\} f features. Firstly, we leverage the capabilities of GPT in capturing previous contextual information and dealing with long-range dependencies in predicting next packet. Evaluating GPT’s predicted next packets given current packets requires an understanding of the context on both sides of the two packets to make predictions about their relationship. Next, exploiting BERT’s bidirectional contextual understanding, we fine-tune it for a packet-pair classification task that aims to predict whether the predicted packet is the next of a given packet. Finally, we train LSTM as a packet classifier, exploiting its proficiency in learning long-term dependencies, which aids in identifying normal and malicious traffic.

%During the deployment phase of our framework (see figure \ref{fig:archi3}) and after collecting and parsing packet, next packets are predicted using packets predictor and classified as normal or malicious using the LSTM packet classifier.

In the deployment phase illustrated in figure \ref{fig:archi3}, after collecting and parsing packets, the packet predictor predicts next packets. The LSTM packet classifier then classifies these packets as either normal or malicious. Notably, this deployment is strategically executed at the Multi-Access Edge Computing (MEC) server level. This positioning enhances the overall effectiveness and responsiveness of our framework.

%Transformer’s \cite{vaswani2017attention} architecture with attention mechanism allows fine-tuned pre-trained large language models to capture long-range relationships and a better contextual comprehension in task-specific data. 

%The proposed framework contains two parts, as illustrated in Figure \ref{fig:archi3}. The first part is two fine-tuned pre-trained large language models, a generative pre-trained (GPT) model to generate next packet given current packet of the same network flow and BERT model for the packet-pair classification task that aims to predict whether one packet is the next of the other, as an evaluation metric for the generative pre-trained (GPT) model. The second part, is the classification of generated packet using  a supervised trained Long Short Term Memory (LSTM) model. The generated next packet can be fed into the generative pre-trained (GPT) model as new current packet, which allows more intrusion prediction in the same network flow.
%thereby reducing the work required to collect and categorize malicious traffic to train the car's communication pattern.

\subsection{Packet parsing and Preparing}
The proposed framework is centered around intrusion prediction based on network packets, focusing on packet headers. Before feeding data into the framework, we initiate the process by calculating a set of packet features for each packet. This involves parsing various application protocols and extracting pertinent information from network packet capture files. These features characterize packet header field values corresponding to the layer 2, 3, and 4 protocol fields of the TCP/IP protocol stack. Notably, a flow index feature facilitates tracking each packet back to its corresponding flow. Concluding the packet parsing phase, we obtain a valid input for our models.\par
\subsection{Pre-trained Large Language Models for Predicting Next Packets} 
In the realm of network traffic, where data is inherently structured as a sequence of packets, we harness the power of transformer architecture \cite{vaswani2017attention}, specifically in sequence-to-sequence tasks, to craft our next packet predictor. This neural network architecture employs the self-attention mechanism, facilitating parallel processing of input sequences and demonstrating remarkable effectiveness in handling sequence-to-sequence tasks. The mathematical expression of this self-attention mechanism in the transformer is as follows: \begin{equation} Attention(Q, K, V ) = softmax \left( \frac{QK^{T}}{\sqrt{d_{k}}} \right )V \end{equation} Here, Q, K, and V represent the query, key, and value matrices, respectively, while $d_k$ signifies the dimensionality of the keys vector. We integrate two LLMs in a feedback loop, fine-tuned GPT model for predicting network traffic and a fine-tuned BERT for evaluating the predicted traffic.\par \subsubsection{Developing an LLM-based next packet predictor} In the initial stage of our Intrusion Prediction framework, we fine-tune the GPT model using a network packet dataset for the purpose of generating high-quality next network packets given current packets while maintaining contextual coherence. GPT operates in a decoder-only configuration within the transformer architecture and pre-trained using a causal language modeling objective (CLM), which means it predicts the next token in a sequence given the preceding context. Initially, the network packet undergoes tokenization as a continuous sequence, breaking it into smaller units or tokens. Subsequently, these tokenized packets are converted into numerical values, accompanied by positional encoding, facilitating GPT's processing while preserving the relative positions of tokens within the input. GPT model applies a multi-headed self-attention operation over the input tokens followed by position-wise feedforward layers to produce an output distribution over target tokens as follows: \begin{equation}\begin{gathered} h_{0} = T W_{e} + W_{p}\\ h_{l} = tf\_block(h_{l-1}) \forall l\in [1,L]\end{gathered}\end{equation} Where T is a matrix of one-hot row vectors of the token indices in the sentence, $W_e$ is the token embedding matrix, and $W_p$ is the position embedding matrix, L is the number of Transformer blocks, and $h_l$ is the state at layer l. This intricate process empowers GPT to capture contextual information and intricate relationships between tokens, considering only previous tokens to generate the next ones. Ultimately, GPT generates a numerical output representing the next network packet based on its comprehensive understanding of the preceding input tokens. Specifically, given an input network packet consisting of N tokens, denoted as P =\{$p_1$ , $p_2$ , . . . , $p_N$\}, GPT calculates the probability $P_N$ of token $t_k$ based on the preceding k-1 tokens: \begin{equation} P_{N}(t_{k}| t_{1},...,t_{k-1})=softmax(W\upsilon h_{k-1}) \end{equation} Where $h_k-1$ denotes the representation encoded by Transformer with the previous tokens \{$t_1$,...,$t_k-1$\} as input. W$\upsilon$ represents the learnable parameters. Thus, the objective is to maximize the likelihood of predicting the next token in a sequence given the preceding context. In mathematical terms, it involves finding the parameters of the model that maximize the probability of the training data: \begin{equation} \mathcal L_{GPT}(\theta)= arg max_\theta\sum_{i=1}^N\operatorname{log}P{(xi\vert x_{<i};\operatorname\theta)}\end{equation}
Here, $\theta$ represents the model parameters, $N$ is the number of tokens in the training dataset, $x_i$ is the $i-th$ token, and $x_{<i}$ represents the context of tokens before $x_i$. The objective is to maximize the log-likelihood of the observed data. This numerical output is subsequently decoded back into the network packet format.\par
\subsubsection{Developing an LLM-based next packet prediction evaluator}
We fine-tune BERT for packet-pair classification task that to evaluate GPT's output. In contrast to GPT, BERT randomly selects a portion of input tokens and replaces them with a "\textit{[MASK]} " token. The model is subsequently trained to predict the original identities of these masked tokens, leveraging contextual information. The objective function is designed to maximize the log-likelihood of predicting the correct tokens within the masked positions and can be expressed as follows:
\begin{equation}
\mathcal L_{BERT}(\theta)= \sum_{i=1}^N\operatorname{log}P{(xi\vert x_{masked};\operatorname\theta)} 
\end{equation}
Here, $N$ is the total number of masked positions, $x_i$ is the true identity of the masked token at position $i$, $x_{masked}$ is the context of the masked tokens, and $\theta$ represents the parameters of the BERT model.\par
Choosing BERT model as an evaluator for GPT’s generated next packets is due to BERT's bidirectional context representation. Considering right and left context is highly effective for the evaluation of the validity of two-packet succession.  Fine-tuning BERT involves preparing a specialized dataset, consisting of labeled pairs of successive and non-successive network packets. In this process, each packet pair (current packet and next packet) undergoes tokenization, conversion into embeddings, and the addition of special tokens for classification and separation. Segment embeddings are incorporated to distinguish between the two packets, with each token assigned a segment embedding indicating its packet of origin. Furthermore, positional embeddings are introduced to convey the positional information of each token in the packet.
Utilizing the Masked Language Model (MLM) technique, a percentage of randomly selected input tokens are intentionally masked. BERT is then tasked with predicting these masked tokens as it traverses through the stack of transformer encoder layers. A classification layer is introduced atop the classification token representation, tailored for the packet-pair classification task. During the fine-tuning process, the model computes the loss by comparing its predictions with the ground truth labels for the packet-pair using a classification loss function. Subsequently, the model's parameters are updated and optimized. During GPT’s  evaluation, the fine-tuned BERT model takes current and generated next packets as input, and the classification layer produces output probabilities, indicating the likelihood of successive or non-successive classes.
\subsection{Training a next packet classifier} We train an LSTM model to classify network packets as normal or malicious as shown in the devlopement phase in Figure \ref{fig:archi3}. Network intrusions often manifest as deviations from normal network behavior. 
The LSTM model consists of two key components: the encoder and the decoder. Working collaboratively, these components aim to learn a compressed representation of the input data and then faithfully reconstruct it. The encoder processes the input sequence, utilizing LSTM cells to compress it into a latent representation. Sequentially processing each element of the input sequence, the encoder captures pertinent information in hidden states. Subsequently, the decoder employs LSTM cells to iteratively generate each element of the reconstructed sequence from the compressed representation generated by the encoder. The operations within an LSTM cell are governed by equations that control the flow of information. With our input vector $x_t$ , the implementation of the LSTM’s cell for the hidden state ht at time t can be represented by the following equations: 
\begin{equation}
\begin{aligned}
i_t &= \sigma(W_{i} x_t + U_{i} h_{t-1} + b_{i}) \\
f_t &= \sigma(W_{f} x_t + U_{f} h_{t-1} + b_{f}) \\
o_t &= \sigma(W_{o} x_t + U_{o} h_{t-1} + b_{o}) \\
g_t &= \pi (W_{g} x_t + U_{g} h_{t-1} + b_{g}) \\
c_t &= f_t \odot c_{t-1} + i_t \odot g_t \\
h_t &= o_t \odot \pi(c_t)
\end{aligned}
\end{equation} 
Here, $\sigma$ denotes the logistic sigmoid function, while $\pi$ represents the non-linear activation functions for cell input and output, typically using the hyperbolic tangent function. Vectors it, $f_t$, $o_t$, $g_t$, and $c_t$ correspond to the input gate, forget gate, output gate, cell input activation, and cell state. The matrices and vectors $W_\alpha$, $U_\alpha$, and $b_\alpha$, where $\alpha$ $\in$ \{i, f, o, g\}, are the parameters to be learned and $\odot$ signifies element-wise product. With their ability to retain and utilize information over extended sequences, LSTM can effectively capture the temporal context of network activities given previously seen packet traffic. This includes recognizing normal patterns and identifying anomalies or suspicious deviations. \par

\section{Performance Evaluation}\label{SIM}

In this section, we initiate with an overview of the dataset utilized in our research. Then,  we pass through the conducted experiments to present used configurations for our models. Finally, we present the obtained results and a discussion of the performance metrics related to our novel intrusion prediction framework.
\subsection{Dataset pre-processing}

We used a realistic IoT attack dataset developed by \cite{s23135941} called CICIoT2023, to fine-tune the pre-trained large language models, GPT and BERT, as well as to train the LSTM model. This dataset, developed in a lab environment with 105 devices, encompasses 33 attacks categorized into DDoS, DoS, Recon, Web-based, brute force, spoofing, and Mirai. We selected five attack types, considering their higher frequency in real-world scenarios. We utilized Tranalyzer \cite{tranalyzer}, a packet exporter to extract packet features from raw network traffic (PCAP files), focusing on layer 2, 3, and 4 protocol fields of the TCP/IP protocol stack. This extraction yielded a total of 71 features, from which 26 were selected after applying feature selection techniques. Constant and quasi-constant features were removed using a minimum Variance Threshold of 25\%. Additionally, highly correlated features $(> 90\%)$ were discarded through a Pearson correlation filter, ensuring a refined set of distinctive and relevant features for analysis. Data is prepared differently for each model.

\subsection{Experiments} We conduct our experiment on Google Colab cloud environment \cite{Bisong2019} using python 3 with the freely available GPU computing to implement the different parts of our intrusion prediction framework.  In our experiment, CICIoT2023 dataset is partitioned between GPT-2, BERT and LSTM models.\par

\subsubsection{Fine-tuning GPT-2 for next packet prediction} To construct fine-tuning dataset for GPT2 model, we transform unlabeled input network packets into textual representation where each line represents a network packet. We add two special tokens that mark the begin and the end of a network flow to the Byte-level BPE (BBPE) Tokenizer (the GPT2’s tokenizer) vocabulary, in order to help the model learn the concept of flow boundaries, thus ensuring prediction of packets of the same network flow and helping GPT-2 model representing and understanding the context and patterns of a network traffic. GPT2’s output has a fixed-length representation of the network packet, since in its input there is the same number of extracted features for each network packets which helps in generating coherent, contextually relevant output and avoiding randomness. We fine-tuned the the small version of GPT-2 model which has 117 million parameters from the HuggingFace transformer Python package\cite{wolf-etal-2020-transformers}.
\par

\subsubsection{Fine-tuning BERT for Packet-Pair Classification} To fine-tune BERT for pair-packet classification task that predict whether one packet is next to another, we constructed a dataset that contains a pair of network packets per each line that represents consecutive or non consecutive class.  This customized dataset is constructed from CICIoT2023 normal and attack network traffic and a binary classification will be performed on it. Available data were split into training, validation, and test sets. We leverage the HuggingFace transformer Python libraries to fine-tune distilbert-base-uncased model \cite{Sanh2019DistilBERTAD}, a small and fast version of the BERT base model. This model has 6 layers, 768 dimension, 12 heads and  66 million parameters. Our model has a sequence classification head on top of its outputs, in order to be able to classify the input pair-packet as positive (non consecutive) or negative (consecutive). An  Early stopping mechanism is  employed to prevent overfitting. The well fine-tuned BERT model is considered as an evaluator of the GPT2’s generated network packets, it classifies the pair of current packets with their corresponding GPT2’s generated next packets. 
\par
\subsubsection{LSTM training for packet classification} The training process of the LSTM packet classifier, utilizing labeled CICIoT2023 network packets, involves several key steps. Initially, we categorize the data into benign or specific attack types, creating a binary classification copy. Table \ref{tab:distSamp} details the distribution of normal and cyber attack samples in the training and evaluation datasets.
\begin{table}[!ht]
\small
\centering
\caption{Dataset samples distribution}
\begin{tabular}{@{}ll@{}}
\toprule
traffic type & instances\\ \midrule
Normal         & 1079391   \\
DDoS       & 55462   \\
Browser Hijacking       & 43414   \\
Command Injection       & 36323   \\
XSS       & 22838   \\
Backdoor Malware       & 19411   \\ \bottomrule
\end{tabular}
\label{tab:distSamp}
\end{table}
To facilitate numerical processing, we convert categorical data into numerical format using an ordinal encoder, and then we normalize the data using a min–max normalization technique following the formula:
\begin{equation}
\begin{aligned}
X_{scaled} = \frac{x-x_{min}}{x_{max}-x_{min}} 
\end{aligned}
\end{equation} where $x$, $X_{scaled}$ represent the original input, and scaled respectively. Similarly, $x_{min}$ and $x_{max}$ are the minimum and the maximum values of the input respectively. After that, we split the data into training, validation, and test sets and reshaping it into three-dimensional input data (number of network packets, size of time step, number of input features)  to ensure the data is appropriately formatted for input into the LSTM. During training, the model learns from the training dataset across multiple epochs, with progress monitored on a validation set. An early stopping mechanism is utilized to prevent overfitting. Post-training, the model undergoes evaluation on the test set, involving the analysis of metrics such as accuracy, precision, recall, and \text{F1-score}. Hyperparameters tuning is carried out to optimize the model's configuration. Refer to Table \ref{tab:hyperparameters} for details on the hyperparameters used in the training process. The trained model is then employed to classify GPT’s predicted network packets. 
\begin{table}[!ht]
\centering
\caption{LSTM's Hyperparameters}
\begin{tabular}{@{}ll@{}}
\toprule
Hyperparameter  & value\\ \midrule
LSTM's neurons       & 64   \\
Optimizer         & Adam   \\
Loss function       & sparse categorical crossentropy   \\
Number of epochs       & 80   \\
Dropout rate       & 0.2   \\
Early stopping patience       & 3   \\
 \bottomrule
\end{tabular}
\label{tab:hyperparameters}
\end{table} \subsection{Results}
To evaluate our framework, we used the following performance metrics: 

\begin{equation}
\text{Accuracy}= \frac{TP+TN}{TP+TN+FP+FN} 
\end{equation}
\begin{equation}
\text{Precision} = \frac{TP}{TP+FP} 
\end{equation}
\begin{equation}
\text{Recall} = \frac{TP}{TP+FN} 
\end{equation}
\begin{equation}
\text{F1-Score} = \frac{2\times Precision \times Recall}{Precision+Recall} 
\end{equation}

where TP (True Positive) denotes the count of instances accurately classified as attacks, while TN (True Negative) signifies the instances correctly identified as normal. Conversely, FP (False Positive) indicates instances incorrectly classified as an attack, and FN (False Negative) represents instances incorrectly classified as normal.\par
Table \ref{tab:PP_perf} presents the obtained results for the packet-pair classification task. With an accuracy of 93.4\%, fine-tuned BERT model demonstrates a high level of overall correctness in predicting whether one packet is next to another. Notably, the model achieves a well-balanced combination of precision and recall, reflected in an impressive F1 score of 96.48\%. These results collectively emphasize the robustness of the fine-tuned BERT model in successfully handling the network packet-pair classification task. 

Due to resource constraints, GPT-2 undergoes fine-tuning over three iterative epochs with only 100,388 instances, representing network packets, including special tokens. This resulted in 20\% of what was generated by GPT-2 being considered correct according to the fine-tuned BERT. These results can be improved by increasing the number of instances dedicated for the fine-tuning of GPT2 and BERT model.

\begin{table}[!ht]
\centering
\caption{Packet-pair classification task results on precision, recall, F1 and accuracy}
\begin{tabular}{@{}llll@{}}
\toprule
Accuracy & Precision & Recall & F1-score\\ \midrule
93.40\%  & 94.07\% & 99.01\% & 96.48\%   \\
\bottomrule
\end{tabular}
\label{tab:PP_perf}
\end{table}

\begin{figure}[ht]
    \centering
    \includegraphics[scale=0.35]{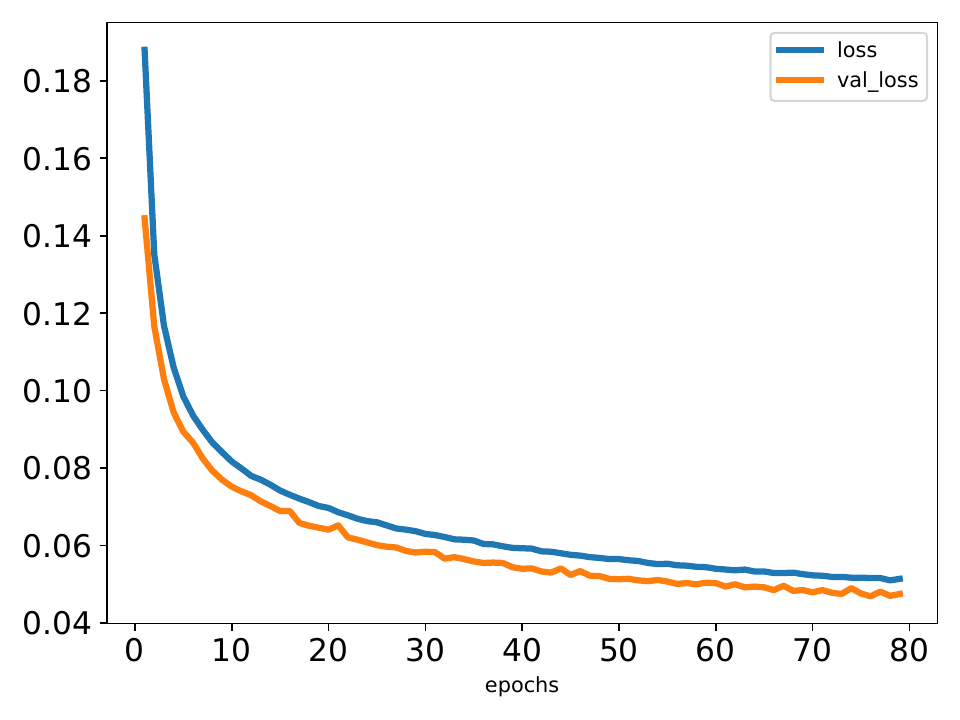}
    \caption{training and validation loss over epochs}
    \label{fig:lstm_loss}
\end{figure}

\begin{figure}[th!]
    \centering
    \includegraphics[width=0.8\columnwidth,height=0.28\textheight]{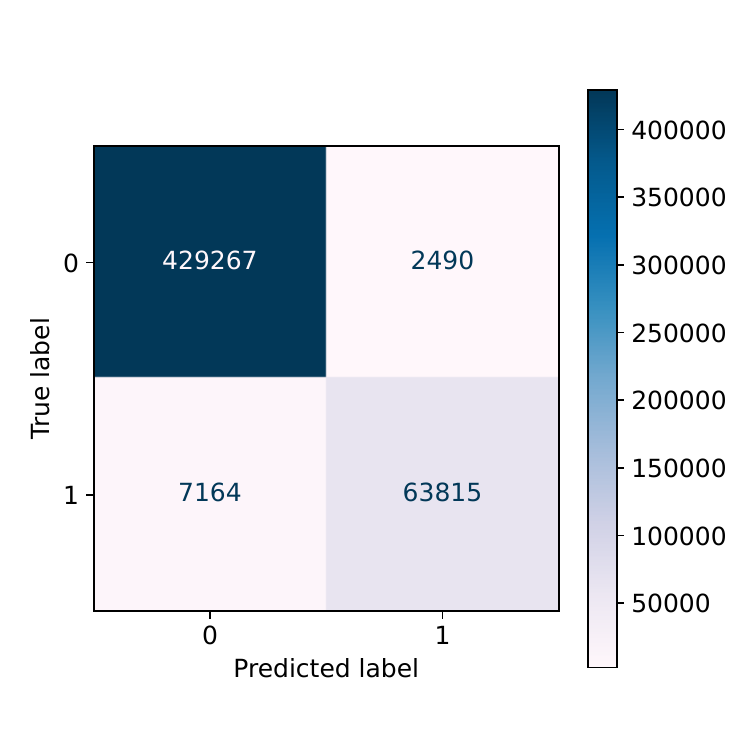}
    \caption{Confusion matrix of LSTM binary classification}
    \label{fig:lstm_cm}
\end{figure}

\begin{figure}[ht!]
    \centering
    \includegraphics[scale=0.35]{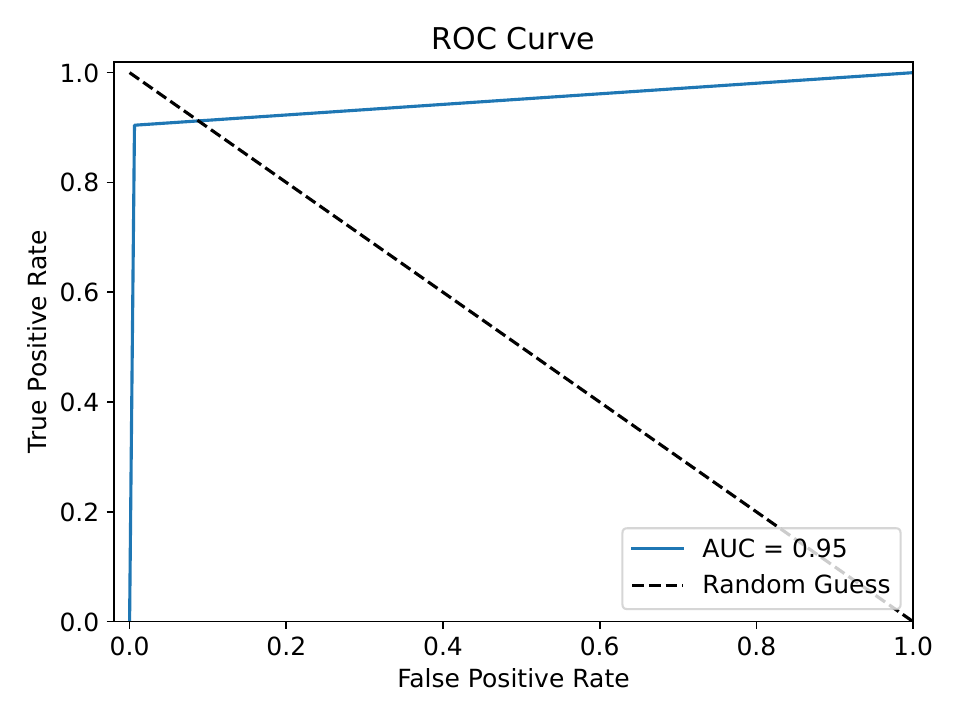}
    \caption{LSTM’s intrusion detection Receiver Operating Curve (ROC)}
    \label{fig:lstm_roc}
\end{figure}

We conducted a binary classification, distinguishing between malicious and normal packets. We can observe from Figure \ref{fig:lstm_loss}  a consistent convergence in loss between the training and validation sets, becoming evident after few tens of epochs. The confusion matrix obtained is depicted in Figure \ref{fig:lstm_cm}.
Figure \ref{fig:lstm_roc} illustrates the Receiver Operating Characteristic (ROC) curve, evaluating the relationship between false positive and true positive rates. The model effectively distinguishes between positive and negative rates, achieving a 95\% separation, as illustrated by the Area Under the Curve (AUC) value.
Table \ref{tab:lstm_report} summarizes the performance outcomes derived from our LSTM binary classification model. The proposed model obtained a testing accuracy of 98\%. The model attained impeccable precision, recall, and F1-score for benign traffic, indicating highly accurate classification performance. It's worth highlighting the high support count for the normal traffic, totaling 431,757 instances. In the case of attack traffic, precision was high, with a marginally lower recall of 0.90, yielding an F1-score of 0.93, reflecting effective intrusion identification performance.

\begin{table}[!ht]
\centering
\caption{Classification report of LSTM intrusion detection model}
\begin{tabular}{@{}lllll@{}}
\toprule
& Precision & Recall & F1-score & support\\ \midrule
Normal  & \textbf{0.98} & \textbf{0.99} & \textbf{0.99}  & 431757 \\
Attack  & 0.96 & 0.90 & 0.93  & 70979 \\
\midrule
Macro Avg  & 0.97 & 0.95 & 0.96  & 502736 \\
Weighted Avg  & 0.98 & 0.98 & 0.98  & 502736 \\
\midrule
Accuracy & \multicolumn{4}{c}{\textbf{0.98}} \\
\bottomrule
\end{tabular}
\label{tab:lstm_report}
\end{table}

\begin{figure}[!ht]
    \centering
    \includegraphics[width=\columnwidth,height=0.25\textheight]{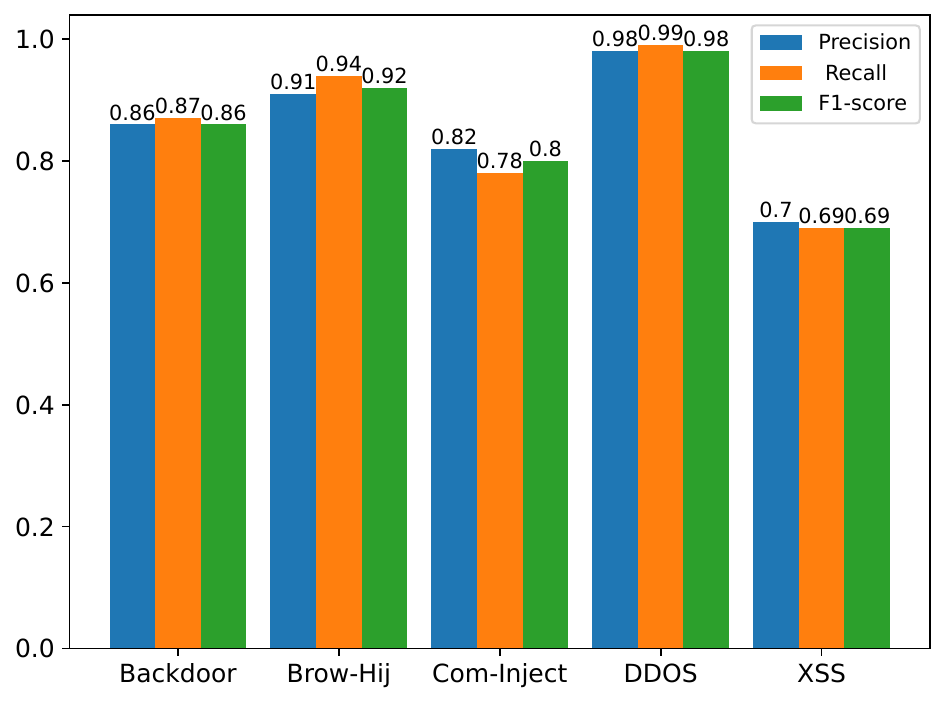}
    \caption{LSTM’s detailed multi-class classification results}
    \label{fig:lstm_attack_classification}
\end{figure}

In addition to the binary classification,
we conducted a multi-class classification involving five attack types (DDoS, BrowserHijacking, CommandInjection, XSS, BackdoorMalware) utilizing LSTM. We can see from Figure \ref{fig:lstm_attack_classification} that most attack types achieved perfect scores in terms of precision, recall, and F1-score, showing a high accurate classification performance on these types. Notably, the XSS attack exhibited lower scores, signifying a misclassification by the model for this particular attack traffic. Addressing this anomaly calls for future work and improvements in the model.

%======================
\section{Conclusion} \label{CON}
The heightened connectivity of IoT devices exposes them to various threats, emphasizing the insufficiency of relying solely on IDSs, as damage often occurs before effective mitigation measures can be applied. Recognizing this, our paper introduces a network packet-based intrusion prediction framework that utilizes GPT, BERT, and LSTM models to enhance IoT security. Relying on fine-tuned GPT and BERT, the proposed framework accurately predicts next network packets, and the incorporation of the LSTM model results in an effective intrusion prediction. This anticipatory approach enables appropriate mitigation procedures before attacks occur. Our experimental findings demonstrate the framework's robust performance, achieving a 98\% accuracy rate even with limited data during fine-tuning. In future work, we aim to assess the proposed intrusion prediction framework using diverse IoT datasets that encompass more sophisticated and contemporary attacks.

%======================
\section*{Acknowledgment}

This work was supported by the 5G-INSIGHT bilateral project (ID: 14891397) / (ANR-20-CE25-0015-16), funded by the Luxembourg National Research Fund (FNR), and by the French National Research Agency (ANR).

%\vspace{12pt}
%\color{red}

\bibliographystyle{unsrt}
\bibliography{ref}

\begin{thebibliography}{10}

\bibitem{abdulganiyu2023systematic}
Oluwadamilare~Harazeem Abdulganiyu, Taha Ait~Tchakoucht, and Yakub~Kayode Saheed.
\newblock A systematic literature review for network intrusion detection system (ids).
\newblock {\em International Journal of Information Security}, pages 1--38, 2023.

\bibitem{heidari2023internet}
Arash Heidari and Mohammad~Ali Jabraeil~Jamali.
\newblock Internet of things intrusion detection systems: A comprehensive review and future directions.
\newblock {\em Cluster Computing}, 26(6):3753--3780, 2023.

\bibitem{saied2024review}
Mohamed Saied, Shawkat Guirguis, and Magda Madbouly.
\newblock Review of artificial intelligence for enhancing intrusion detection in the internet of things.
\newblock {\em Engineering Applications of Artificial Intelligence}, 127:107231, 2024.

\bibitem{asharf2020review}
Javed Asharf, Nour Moustafa, Hasnat Khurshid, Essam Debie, Waqas Haider, and Abdul Wahab.
\newblock A review of intrusion detection systems using machine and deep learning in internet of things: Challenges, solutions and future directions.
\newblock {\em Electronics}, 9(7):1177, 2020.

\bibitem{Liu2024}
Junming Liu, Yuntao Zhao, Yongxin Feng, Yutao Hu, and Xiangyu Ma.
\newblock Semalbert: Semantic-based malware detection with bidirectional encoder representations from transformers.
\newblock {\em Journal of Information Security and Applications}, 80:103690, 2 2024.

\bibitem{Seyyar2022}
Yunus~Emre Seyyar, Ali~Gökhan Yavuz, and Halil~Murat Ünver.
\newblock An attack detection framework based on bert and deep learning.
\newblock {\em IEEE Access}, 10:68633--68644, 2022.

\bibitem{Chen2022}
Song Chen and Hai Liao.
\newblock Bert-log: Anomaly detection for system logs based on pre-trained language model.
\newblock {\em Applied Artificial Intelligence}, 36, 12 2022.

\bibitem{Ferrag2023}
Mohamed~Amine Ferrag, Mthandazo Ndhlovu, Norbert Tihanyi, Lucas~C Cordeiro, Merouane Debbah, Thierry Lestable, and Narinderjit~Singh Thandi.
\newblock Revolutionizing cyber threat detection with large language models: A privacy-preserving bert-based lightweight model for iot/iiot devices.
\newblock {\em IEEE Access}, 2024.

\bibitem{Ferrag2023SecureFalcon}
Mohamed~Amine Ferrag, Ammar Battah, Norbert Tihanyi, Merouane Debbah, Thierry Lestable, and Lucas~C. Cordeiro.
\newblock Securefalcon: The next cyber reasoning system for cyber security.
\newblock {\em arXiv preprint arXiv:2307.06616}, 7 2023.

\bibitem{han2023loggpt}
Xiao Han, Shuhan Yuan, and Mohamed Trabelsi.
\newblock Loggpt: Log anomaly detection via gpt.
\newblock In {\em 2023 IEEE International Conference on Big Data (BigData)}, pages 1117--1122. IEEE, 2023.

\bibitem{radford2019language}
Alec Radford, Jeffrey Wu, Rewon Child, David Luan, Dario Amodei, Ilya Sutskever, et~al.
\newblock Language models are unsupervised multitask learners.
\newblock {\em OpenAI blog}, 1(8):9, 2019.

\bibitem{devlin2019bert}
Jacob Devlin, Ming-Wei Chang, Kenton Lee, and Kristina Toutanova.
\newblock Bert: Pre-training of deep bidirectional transformers for language understanding.
\newblock In {\em Proceedings of the 2019 Conference of the North American Chapter of the Association for Computational Linguistics: Human Language Technologies, Volume 1 (Long and Short Papers)}, pages 4171--4186, 2019.

\bibitem{Wang2024}
Zhendong Wang, Jingfei Li, Shuxin Yang, Xiao Luo, Dahai Li, and Soroosh Mahmoodi.
\newblock A lightweight iot intrusion detection model based on improved bert-of-theseus.
\newblock {\em Expert Systems with Applications}, 238:122045, 3 2024.

\bibitem{penedo2024refinedweb}
Guilherme Penedo, Quentin Malartic, Daniel Hesslow, Ruxandra Cojocaru, Hamza Alobeidli, Alessandro Cappelli, Baptiste Pannier, Ebtesam Almazrouei, and Julien Launay.
\newblock The refinedweb dataset for falcon llm: Outperforming curated corpora with web data only.
\newblock {\em Advances in Neural Information Processing Systems}, 36, 2024.

\bibitem{omar2023vuldetect}
Marwan Omar and Stavros Shiaeles.
\newblock Vuldetect: A novel technique for detecting software vulnerabilities using language models.
\newblock In {\em 2023 IEEE International Conference on Cyber Security and Resilience (CSR)}, pages 105--110. IEEE, 2023.

\bibitem{vaswani2017attention}
Ashish Vaswani, Noam Shazeer, Niki Parmar, Jakob Uszkoreit, Llion Jones, Aidan~N Gomez, {\L}ukasz Kaiser, and Illia Polosukhin.
\newblock Attention is all you need.
\newblock {\em Advances in neural information processing systems}, 30, 2017.

\bibitem{s23135941}
Euclides Carlos~Pinto Neto, Sajjad Dadkhah, Raphael Ferreira, Alireza Zohourian, Rongxing Lu, and Ali~A. Ghorbani.
\newblock Ciciot2023: A real-time dataset and benchmark for large-scale attacks in iot environment.
\newblock {\em Sensors}, 23(13), 2023.

\bibitem{tranalyzer}
S.~{Burschka} and B.~{Dupasquier}.
\newblock Tranalyzer: Versatile high performance network traffic analyser.
\newblock In {\em 2016 IEEE Symposium Series on Computational Intelligence (SSCI)}, pages 1--8, 2016.

\bibitem{Bisong2019}
Ekaba Bisong.
\newblock {\em Google Colaboratory}, pages 59--64.
\newblock Apress, Berkeley, CA, 2019.

\bibitem{wolf-etal-2020-transformers}
Thomas Wolf, Lysandre Debut, Victor Sanh, Julien Chaumond, Clement Delangue, Anthony Moi, Pierric Cistac, Tim Rault, Remi Louf, Morgan Funtowicz, Joe Davison, Sam Shleifer, Patrick von Platen, Clara Ma, Yacine Jernite, Julien Plu, Canwen Xu, Teven Le~Scao, Sylvain Gugger, Mariama Drame, Quentin Lhoest, and Alexander Rush.
\newblock Transformers: State-of-the-art natural language processing.
\newblock In Qun Liu and David Schlangen, editors, {\em Proceedings of the 2020 Conference on Empirical Methods in Natural Language Processing: System Demonstrations}, pages 38--45, Online, October 2020. Association for Computational Linguistics.

\bibitem{Sanh2019DistilBERTAD}
Victor Sanh, Lysandre Debut, Julien Chaumond, and Thomas Wolf.
\newblock Distilbert, a distilled version of bert: smaller, faster, cheaper and lighter.
\newblock {\em ArXiv}, abs/1910.01108, 2019.

\end{thebibliography}

\end{document}